# Algorithms for Locating and Characterizing Cable Faults via Stepped-Frequency Waveform Reflectometry

Nicola Giaquinto, *Member, IEEE,* Marco Scarpetta, *Student Member, IEEE,* and Maurizio Spadavecchia, *Member, IEEE*

Polytechnic University of Bari

*Abstract*—The paper presents algorithms to realize effectively and accurately the stepped-frequency waveform reflectometry (SFWR), i.e. the reflectometric technique based on the use of sinusoidal bursts. This technique is useful for monitoring the health status of connection cables, but has many other applications, like other reflectometric techniques. The paper outlines the theory of SFWR, highlighting the problems associated to the transient components in the reflected signals; presents a method to minimize the effect of the transients, estimating the frequency response function (FRF) of interest with very low systematic error; shows how to use the FRF to locate and characterize faults in cables; evaluates accurately, using simulated cables with exactly known characteristics, the errors associated to the proposed methods. Overall, the paper demonstrates how the SFWR technique can be effectively used for testing cables, and in general determine, via reflectometry, parameters of interest of transmission lines.

## I. Introduction

Reflectometry is a technique proposed in 1931 [1], and used since the Forties to detect and localize defect in cables [2]. It is nowadays used for a number of applications in very different fields, e.g.: for measuring humidity and salinity of media and materials (like soil, concrete, etc.) [3], [4], [5]; for monitoring landslide and rock movement [6], [7]; for testing circuits and PCB [8], [9]; for measuring the level of liquids [10] [11], [12]; for leak detection in underground water pipes [13], [14]; etc.

Different kind of signals can be used, and therefore, different kind of reflectometric methods are available. Comparisons among different methods are presented in [15], in [16], and in [17] (where the new method of impedance analysis at the cable input is suggested). The most used reflectometric methods are the following:

1. Time-domain reflectometry (TDR): it is probably the most common, and uses narrow pulses or short rise time steps;
2. Frequency domain reflectometry (FDR): it uses stepped-frequency sinusoids, and is actually a family of techniques, mainly phase-detection FDR (PD-FDR), standing wave reflectometry (SWR), and frequency-modulated continuous-wave reflectometry (FMCW-FDR);
3. Time-frequency domain reflectometry: uses Gaussian chirp signals [18];
4. Sequence TDR and Spread Spectrum TDR: use, respectively, baseband and passband pseudorandom signals [19], [20], [21].

In a recent paper [22], a new reflectometric technique has been proposed. The technique, named stepped-frequency waveform reflectometry (SFWR) uses sinusoidal bursts, and therefore combines some advantages inherent of FDR and TDR. First, generating sinusoidal bursts does not require a fast rise-time pulse generator like TDR, nor a swept sinusoidal generator like FMCW-FDR. Second, the analysis of sinusoidal bursts does not require special hardware with directional couplers to separate the reflected and the transmitted signal, like in PD-FDR. Third, sinusoidal bursts can be either embedded in a single signal to obtain the complete measurement with a single acquisition, or, if more convenient, generated and used once at a time with a cheap sinusoidal generator; the measurement is carried out with multiple acquisition, without any loss of accuracy. Summing up, using sinusoidal bursts makes possible accurate frequency-domain measurements with cheap and portable hardware. Furthermore, SFWR is compared with TDR in [22], using practical experiments. Results show that an accurate assessment of faults characteristics (location, reflection coefficient) is difficult to obtain with TDR, and they are influenced by the distance of the fault. It is true, on the other hand, that TDR can be improved with dedicated signal processing and system identification techniques, like those proposed in [24]. A comprehensive comparison between SFWR and TDR could be the subject of a separate paper.

In [22], the effectiveness of SFWR is demonstrated using specific signal processing techniques, and under specific assumptions. As regards signal processing, the work relies on the Rihaczek time–frequency distribution, a non-parametric technique that is able to obtain good results, but does not exploit additional information available on the signal. In particular, it can be usefully taken into account that the reflected signal contains transient components which affect a non-parametric frequency analysis. As regards the underlying assumptions, all measurements in [22] are obtained assuming a quadratic model of the propagation function, and a frequency-independent reflection coefficient. It is obviously desirable to release these assumptions. Moreover, although the experimental results presented in [22] are clearly satisfactory, only the order of magnitude of the measurement errors can be evaluated, by comparison with values assumed as reference (e.g. the propagation function measured with open-ended cable). Simulation results, with exactly known "true values" of the parameters to be measured, are desirable in order to evaluate precisely the performance of the estimation algorithms.

The aim of the paper is to present:
- a comprehensive theoretical framework of the SFWR technique, with a clear systematization of the basis of the method, built on a more general propagation model;
- signal estimation algorithms designed specifically to work with sinusoidal bursts reflected by a transmission line;
- measurement techniques free from assumptions about the propagation function and the reflection coefficient;
- the optimization of the above techniques for the special case of reflection coefficient independent on the frequency, examined in [22];
- simulation results, using a MATLAB simulator already developed and validated by the authors [23], in order to evaluate accurately the errors due to the algorithms;
- experimental results, obtained from measurements on real cables, to prove the effective applicability of proposed algorithms in real-life situations.

Therefore, estimation algorithms are introduced in this paper, different from those presented in [22], and able to better extract relevant information from SFWR signals. Furthermore, the technique has been extended to work in more general situations, so that limiting hypothesis about the propagation function of the line and the reflection coefficient are no more necessary.

Even if reflectometry has many different applications, as pointed out above, the paper is focused on the location and classification of faults in cables. This is to keep the focus on a specific problem, and for easy comparison with [22]. Of course, monitoring the health status of cables is also very important per se, to guarantee safe and correct operation of many systems and plants, especially those where a failure can cause disastrous accidents, like nuclear power plants and aircrafts.

## II. PRELIMINARY CONCEPTS

In this section we illustrate the basic concepts of SFWR to localize and characterize faults in cables.

### A. Signal reflection and associated FRF in a cable

A cable is a transmission line (TL) which, unless perfectly homogeneous and terminated on its characteristic impedance, has one or more impedance discontinuities, where reflections occurs. We can consider the simple situation in Fig. 1, where a TL with characteristic impedance $Z_0$ is terminated on the load $Z_L$.

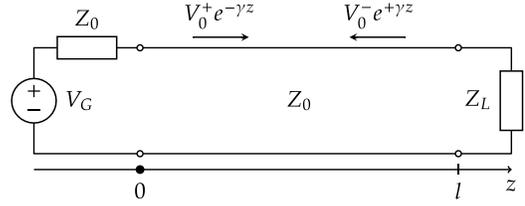

Fig. 1 Model of a cable as a transmission line terminated on a generic load.

The propagation function of the line is:

$$\gamma(\omega) = \sqrt{[R(\omega) + j\omega L(\omega)] \cdot [G(\omega) + j\omega C(\omega)]} = \alpha(\omega) + j\beta(\omega) \quad (1)$$

where $R(\omega), L(\omega), G(\omega), C(\omega)$ are respectively resistance, inductance, conductance, capacitance per unit length (primary parameters), and $\alpha(\omega)$ and $\beta(\omega)$ are the attenuation and propagation functions (secondary parameters). In common idealized models, like the low-loss TL with frequency-independent primary parameters, $\alpha(\omega)$ is a constant, and $\beta(\omega)$ is a linear function. In real-world case, $\gamma(\omega)$ is a more complex function of the frequency.

The phasor of a sinusoidal signal in the TL is:

$$V(z) = V^+(z) + V^-(z) = V_0^+ e^{-\gamma z} + V_0^- e^{\gamma z} \quad (2)$$

where $V^+(z)$ and $V^-(z)$ are, respectively, the phasors of the transmitted and of the reflected signal. The ratio of the phasors at the beginning of the cable ($z = 0$) is a frequency response function (FRF):

$$\bar{H}(\omega) = \frac{V^-(0)}{V^+(0)} = \frac{V_0^-}{V_0^+} = \bar{\Gamma}(\omega) e^{-\gamma(\omega)2l} \quad (3)$$

where $\bar{\Gamma}(\omega) = (Z_L - Z_0)/(Z_L + Z_0)$ is the complex reflection coefficient at the termination of the cable, with magnitude $\Gamma(\omega)$ and phase $\varphi_\Gamma(\omega)$, $l$ is the length of the TL. If the generator is not exactly matched with the TL, (3) is perfectly valid, but $\bar{\Gamma}(\omega)$ includes a term due to the mismatch, as demonstrated in [24].

Amplitude and phase of $\bar{H}(\omega)$ are:

$$H(\omega) = \Gamma(\omega) e^{-\alpha(\omega)2l} \quad (4)$$

$$\varphi_H(\omega) = -\beta(\omega)2l + \varphi_\Gamma(\omega) \quad (5)$$

It is useful to write the phase response also in terms of the propagation time $\tau_p(\omega) = 2l/v_p(\omega)$, where $v_p(\omega) = \omega/\beta(\omega)$ is the propagation velocity:

$$\varphi_H(\omega) = -\omega\tau_p(\omega) + \varphi_\Gamma(\omega) \quad (6)$$

The SFWR technique is primarily designed to measure $\bar{H}(\omega)$ in a set of chosen frequencies $\omega_i$, $i = 0: N - 1$. From $\bar{H}(\omega)$, using other available information, quantities of interest can be obtained, mainly the position $l$ at which the reflection occurs, the propagation functions $\alpha(\omega)$, $\beta(\omega)$, and the reflection coefficient $\bar{\Gamma}(\omega)$.

*B. SFWR transmitted signal, bursts duration and spacing*

The SFWR transmitted signal is a sequence of sinusoidal bursts of increasing, linearly spaced frequencies $\omega_i = \omega_0 + i \cdot \Delta\omega$, $i = 0, \dots, N-1$. The overall signal is $x_{tr}(t) = \sum_{i=0}^{N-1} x_{tr,i}(t)$, with

$$x_{tr,i}(t) = d_{tr,i} \sin(\omega_i(t - iT) + \varphi_{tr,i}) \cdot \text{rect}\left(\frac{t - iT - \frac{\tau}{2}}{\tau}\right) \quad (7)$$

where the amplitudes $d_{tr,i}$ are nominally equal, the phases $\varphi_{tr,i}$ are nominally zero, and $\text{rect}(t)$ is the rectangular unit pulse. An example of such a signal is depicted in Fig. 2.

The critical parameters to choose are:
- the burst duration, $\tau$;
- the time interval between consecutive bursts, $T$;
- the linearly spaced frequencies $\omega_i$.

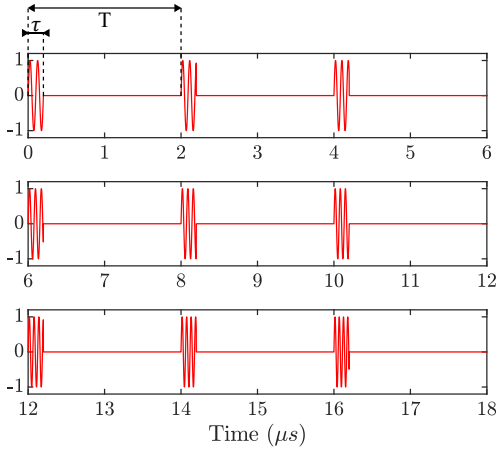

Fig. 2 Example of SFWR transmitted signal.

Discussing the choice of the frequencies needs some specific results, that will be derived in the next Section. As regards $\tau$ and $T$, they must be chosen on the basis of an approximate knowledge of the propagation velocity $v_p$ in the cable, and of the diagnostic range, i.e. the minimum and the maximum positions $[l_{min}, l_{max}]$ at which the reflection can occur. Indeed, the reflection of each burst must satisfy two constraints: *(i)* it begins after the end of the transmitted burst; *(ii)* it vanishes before the beginning of the subsequent transmitted burst.

The first constraint gives the condition:

$$\tau \leq \tau_{max} = \frac{2\, l_{min}}{v_p} \quad (8)$$

and the second constraint gives the condition:

$$T \geq T_{min} = \frac{2\, l_{max}}{v_p} + \tau \quad (9)$$

As an example of design, in the tests reported in Section V, we have $v_p \cong 2 \cdot 10^8$ m/s, $[l_{min}, l_{max}] \cong [20\text{ m}, 180\text{ m}]$, and therefore, $\tau_{max} = 200$ ns. Assuming $\tau = \tau_{max}$, condition (9) gives $T_{min} = 2$ μs.

## III. MEASURING FREQUENCY RESPONSE USING SFWR

As stated in the Introduction, we use time-domain and parametric estimation techniques to analyze the SFWR signal. The advantage is that they can be easily tailored to the actual FRF to be measured. The response of $\bar{H}(\omega)$ to a finite-duration sinusoidal burst, indeed, contains transient components which makes it quite different from the nearly ideal signal depicted in Fig. 2.

*A. Preliminary operations on the SFWR signal*

First of all, the SFWR signal must be split into $N$ segments of duration $T$, so that the $i$-th segment contains a single pair of transmitted and reflected bursts at frequency $\omega_i$. From each segment two time-sequences, respectively containing the transmitted and the reflected burst only, are easily generated, using the knowledge of the parameter $\tau$ (and assuming that the requirements in Subsection II.B are met). The obtained sequences, denoted with $x_n$ and $y_n$, are the input and the output of $\bar{H}(\omega)$: an example is shown in Fig. 3. Both, conventionally, sampled at instants $t_n = nT_s$, where $T_s$ is the sampling interval and $n = 0, 1, \dots, N_s - 1$.

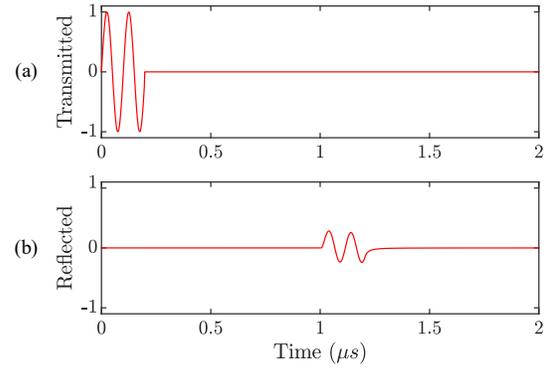

Fig. 3 Example of signals $x_n$ and $y_n$ containing, respectively, the transmitted (a) and the reflected (b) burst extracted by a single segment of duration $T$.

*B. Raw estimation of the time delay $\tau_d$ between $x_n$ and $y_n$*

The time delay between $x_n$ and $y_n$ is, by definition:

$$\tau_d(\omega_i) = -\frac{\varphi_H(\omega_i)}{\omega_i} \quad (10)$$

It is important to note that, in general, $\tau_d = \tau_p(\omega_i) - \varphi_\Gamma(\omega_i)/\omega_i \neq \tau_p$. A raw estimate of $\tau_d$ is easily obtained by computing the cross-correlation $r_n = \text{xcorr}(y_n, x_n)$: if $m$ is the integer lag maximizing the cross-correlation, the estimate is:

$$\tau_{d,raw}(\omega_i) = m \cdot T_s \tag{11}$$

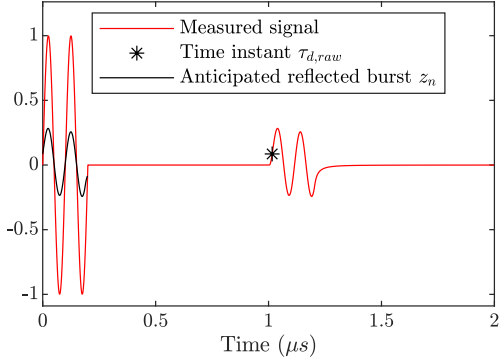

Fig. 4 Illustration of the operation of anticipating the reflected burst of the time $\tau_{d,raw}$, obtaining the signal $z_n$.

*C. FRF estimation via modified sinusoidal fitting*

   *1) Outline of the method*

The general methodology to estimate $H(\omega_i)$ and $\varphi_H(\omega_i)$ is the following, and relies on ordinary least squares (OLS) fitting.

1. Compute $\tau_{d,raw} = m \cdot T_s$.
2. Compute the sequence $z_n = y_{n-m}$, i.e., $y_n$ anticipated of the time $\tau_{d,raw}$ (Fig. 4).
3. Determine the estimated amplitude and phase $\hat{d}_{tr,i}, \hat{\varphi}_{tr,i}$ by OLS fitting on the input sequence $x_n$.
4. Determine the estimated amplitude and phase $\hat{d}_{re,i}, \hat{\varphi}'_{re,i}$ by OLS fitting on anticipated output sequence $z_n$.
5. Compute the estimated phase of the sequence $y_n$, given by $\hat{\varphi}_{re,i} = \hat{\varphi}'_{re,i} - \omega_i \tau_{d,raw}$
6. Compute $\widehat{H}(\omega_i) = \frac{\hat{d}_{re,i}}{\hat{d}_{tr,i}}$, $\hat{\varphi}_H(\omega_i) = \hat{\varphi}_{re,i} - \hat{\varphi}_{tr,i}$

The preliminary shifting of $y_n$ of the quantity $\tau_{d,raw}$ assures that the estimate $\hat{\varphi}_{re,i}$ is not "wrapped" in the interval $[0; 2\pi[$. As a consequence, it is possible to obtain from $\hat{\varphi}_H(\omega_i)$ a correct refined estimation of the time delay using (10), i.e. $\hat{\tau}_d = -\hat{\varphi}_H(\omega_i)/\omega_i$.

To determine $\hat{d}_{tr,i}, \hat{\varphi}_{tr,i}$ it is sufficient a standard OLS sinusoidal fitting on the sequence $x_n$ in the interval $[0; \tau[$. The same fitting is possible on the sequence $z_n$, but the results would be affected by model errors. Fig. 3b shows quite clearly that the sinusoidal model for the reflected burst is not a very good model, and a modified fitting is desirable.

   *2) Modified sinusoidal fitting on $z_n$*

Since in real-world TL $\bar{H}(\omega)$ is almost always low-pass, we must examine the response of a low-pass filter to the finite-duration sinusoidal burst (7), anticipated of the time $iT$. To gain some insight in the problem, we consider the elementary first-order lowpass filter

$$\bar{H}_{LP}(j\omega) = \frac{1}{1 + j\omega\tau_c} \tag{12}$$

so that the response has a tractable analytic expression. This first-order filter is a simplification of the FRF of real cables and its time-domain response to sinusoidal bursts doesn't contain a delay term (like sequence $z_n$ defined above). With the symbols $H_{LP} = |\bar{H}_{LP}(j\omega_i)|$ and $\varphi_{LP} = \angle \bar{H}_{LP}(j\omega_i)$, the response $z(t)$ is the sum of three terms:

$$\begin{aligned} z(t) = & H_{LP} d_{tr,i} \sin(\omega_i t + \varphi_{tr,i} + \varphi_{LP}) \text{rect}\left(\frac{t - \frac{\tau}{2}}{\tau}\right) + \\ & + e^{-t/\tau_c} \frac{\omega_i \omega_c}{\omega_i^2 + \omega_c^2} u(t) + \\ & + e^{-\omega_c(t-\tau)} H_{LP} d_{tr,i} \sin(\omega_i \tau + \varphi_{tr,i} + \varphi_{LP}) u(t - \tau) \end{aligned} \tag{13}$$

where $u(t)$ is the Heaviside step function. The first-order filter response (13) can be easily computed in the Laplace s-domain as:

$$z(t) = \mathcal{L}^{-1}\left\{H_{LP}(s) \cdot \mathcal{L}\{x_{tr,i}(t + iT)\}\right\}$$

The three terms are, respectively:

1) the steady-state response of (12) to $\sin(\omega_i t)$, in the interval $[0; \tau[$;
2) a decreasing exponential term, starting at $t = 0$;
3) a decreasing exponential term, starting at $t = \tau$.

The first term is the one of interest, while the second and the third are transients that jeopardize the estimation if a simple sinusoidal fitting on the whole response is used. Fig. 5 shows examples of the response (13) for different values of the ratio $\tau_c/\tau$. Both the first term and the total response are plotted.

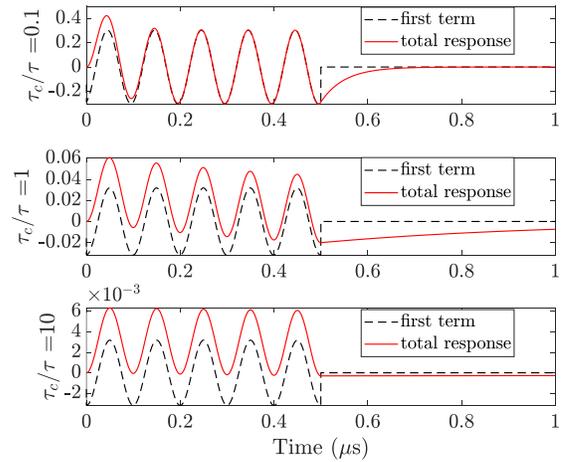

Fig. 5 Examples of response to a sinusoidal burst of a first-order lowpass filter.

Formula (13) and Fig. 5 show clearly that, in order to estimate the parameters of interest $H_{LP}, \varphi_{LP}$ without knowing the value of $\tau_c/\tau$, it is necessary a modified procedure, effectively independent on the ratio $\tau_c/\tau$. This goal is achieved with two devices:

a) consider only *the second half* of the interval $[0, \tau[$, so that a possible initial spike caused by a "fast" exponential term is discarded (first case $\tau_c/\tau \ll 1$ in Fig. 5);
b) fit a sinusoid *with a constant term and a linear trend*, in order to model, even if with some approximation, the possible presence of a "slow" exponential term (second and third case in Fig. 5).

These two devices are both always used in the estimation procedure, independently on the (unknown) $\tau_c/\tau$ ratio, in order to minimize errors due to the exponential term: when $\tau_c/\tau \ll 1$ the exponential terms adds an initial overshoot to the sinusoid, while, when the time constant $\tau_c$ is larger, its effect is modeled with the constant and linear term in the second half of the burst. The model to fit is therefore:

$$z(t) = c + m\,t + d_{tr,i} \sin(\omega_i t + \varphi'_{tr,i}) \quad (14)$$

where the estimated terms $\hat{c}$, $\hat{m}$ are discarded, while $\hat{d}_{tr,i}$ and $\hat{\varphi}'_{tr,i}$ are used to estimate the FRF.

The performance of the approximate model (14) have been evaluated for different values of $\tau_c/\tau$, obtained varying $\tau_c$ while keeping $\tau$ constant. The estimation has been performed on noiseless signals, since the aim is quantifying only model errors. Results are summarized in Table I that reports, for each case, the relative magnitude estimation error $e_r(d_{tr,i}) = (\hat{d}_{tr,i} - d_{tr,i})/d_{tr,i}$, and the absolute phase estimation error $e(\varphi'_{tr,i}) = \hat{\varphi}'_{tr,i} - \varphi'_{tr,i}$. Model errors are very small, and in most practical applications will be negligible with respect to those introduced e.g. by noise.

TABLE I
MAGNITUDE AND PHASE ESTIMATION ERRORS

| $\tau_c/\tau$ | $e_r(d_{tr,i})$ | $e_r(\varphi'_{tr,i})$ (deg) |
|---|---|---|
| 0.01 | $-7.43 \times 10^{-11}$ | $-2.10 \times 10^{-8}$ |
| 0.1 | $-8.31 \times 10^{-6}$ | 0.0107 |
| 1 | $-4.19 \times 10^{-6}$ | 0.0598 |
| 10 | $-6.47 \times 10^{-9}$ | 0.000842 |
| 100 | $8.15 \times 10^{-11}$ | $8.62 \times 10^{-6}$ |

Model (14) has been derived and evaluated for the simple filter (12), but its effectiveness is furtherly demonstrated by other computer simulations on realistic cables and filters, reported in Section V. In this case, the fitting has been applied to the signal in the interval $[0.5\tau; 0.95\tau]$ (specified in terms of time instants for the sequence $z_n$). The signal in $[0.95\tau; \tau]$ is discarded to avoid possible problems due to the imperfect estimate of $\tau_{d,raw}(\omega_i)$ which could lead to the inclusion, in the fitting interval, of regions where the burst is terminated.

Since the fitting uses only the second half of the signal, it is recommended that the burst at the minimum frequency $\omega_0$ has at least two periods, so that one period is used to reconstruct the signal amplitude and phase. A criterion for the choice of this frequency follows:

$$\omega_0 \geq \frac{4\pi}{\tau} \Leftrightarrow f_0 \geq \frac{2}{\tau} \quad (15)$$

For example, if $\tau = 200$ ns, the requirement is $f_0 \geq 10$ MHz.

## IV. FAULT LOCATION AND CHARACTERIZATION FROM FRF

Having measured $H(\omega_i)$ and $\varphi_H(\omega_i)$, it is possible to perform essentially two kind of measurements:

1. the preliminary characterization of a *reference cable*, of known length and with known termination;
2. the fault location and characterization on a *cable under test*, of the same kind of the characterized reference cable.

Fault location and characterization is, in general, more precise if some information is known about the fault. We examine the case of a generic fault, for which no specific prior information is available, and that (examined also in [22]) of a fault with reflection coefficient constant in the frequencies $\omega_i$ of interest ($\bar{\Gamma}(\omega_i) = \bar{\Gamma}$).

In the following, the fault location $l$ is estimated taking always into account that the propagation time $\tau_p(\omega)$ depends on the frequency. However, the estimated $\bar{H}(\omega_i)$ can be used also to identify an approximate model with a frequency-independent delay, like that in [24] (where $\bar{H}$ is approximated by a rational transfer function cascaded with a delay) or that in [26] (where simple criteria for cost-effective monitoring via TDR are compared).

### A. Characterization of a reference cable

In this case, the cable length $l$ and the reflection at the cable termination, $\bar{\Gamma}(\omega)$, are supposed to be known, and the quantity of interest is the propagation function $\gamma(\omega) = \alpha(\omega) + j\beta(\omega)$. Relations (4) and (5) solves the problem immediately. In the common case of open-ended reference cable ($\bar{\Gamma} = 1$) we have:

$$\hat{\alpha}(\omega_i) = -\frac{\ln \hat{H}(\omega_i)}{2\,l} \quad (16)$$

$$\hat{\beta}(\omega_i) = -\frac{\hat{\varphi}_H(\omega_i)}{2\,l} \quad (17)$$

### B. Location and characterization of a generic fault

In this case, $\gamma(\omega)$ is known from a previous characterization of the cable, and the quantities of interest are the fault location $l$, and the reflection coefficient $\bar{\Gamma}(\omega)$ at the frequencies $\omega_i$. Equations (4) and (5), written for the $N$ frequencies $\omega_i$, are a system of $2N$ equations with the $2N + 1$ unknowns $l$, $\Gamma(\omega_i)$

$\varphi_\Gamma(\omega_i)$. Therefore, a unique solution can be obtained only with further information, e.g. a constraint on the values of $\bar\Gamma(\omega_i)$. Such a case is examined in the next Subsection.

With no further information, it is only possible, under the reasonable assumption of bounded $\varphi_\Gamma(\omega)$, to measure:
- $l$, with a bounded systematic error,
- $\Gamma(\omega_i)$, with a systematic error determined by that on $l$.

The measurement of $l$ is based on the approximation $\tau_p \cong \tau_d$, i.e.:

$$\varphi_H(\omega) = -\omega\tau_d(\omega) = -\omega\tau_p(\omega) + \varphi_\Gamma(\omega) \cong -\omega\tau_p(\omega) \qquad (18)$$

The estimation of $l$ is, consequently, from (5):

$$\hat l = -\frac{\hat\varphi_H(\omega)}{2\hat\beta(\omega)} \qquad (19)$$

The approximation $\tau_p \cong \tau_d$ introduces a relative systematic error, given by:

$$e_r = \frac{\hat l - l}{l} = \frac{\tau_d - \tau_p}{\tau_p} = -\frac{\varphi_\Gamma(\omega)}{\omega_i \tau_p(\omega)} = -\varphi_\Gamma(\omega)\frac{v_p(\omega)}{2l\omega} \qquad (20)$$

Assuming $|\varphi_\Gamma(\omega)| \le \varphi_{\Gamma\max}$ and $v_p(\omega_i) \le v_{pmax}$, a bounded systematic error $|e_r| \le U_r$ is achieved, under the condition

$$\omega \ge \varphi_{\Gamma\max}\frac{v_{pmax}}{2l \cdot U_r} \qquad (21)$$

From this inequality, a criterion to choose the maximum frequency $\omega_{max} = \omega_{N-1}$ arises. Actually, equation (21) only suggests that at least one frequency must satisfy the bound, but, in practice, there is no advantage if more frequencies do, since $l$ is computed using the maximum frequency only.

For example, if $U_r = 1\%$, $l = 180$ m, $\max|\varphi_\Gamma(\omega)| = 2\pi$, and $v_p \cong 2 \cdot 10^8$ m/s, the highest frequency should be at least $f_{N-1} = \omega_{N-1}/2\pi \cong 55$ MHz. With respect to the method for estimating $l$ used in [22] (time-range domain transform), (19) does not require many measurements at different frequencies, but only one measurement at the highest possible frequency. Both methods assume $\tau_d \cong \tau_p$, and are therefore prone to the same systematic error.

As regards the estimation of the reflection coefficient, relation (4) gives:

$$\hat\Gamma(\omega_i) = \hat H(\omega_i)\,e^{2\hat\alpha(\omega_i)\hat l} \qquad (22)$$

Phase $\varphi_\Gamma(\omega_i)$ cannot be estimated using (5), because the approximation $\tau_p(\omega_{N-1}) \cong \tau_d(\omega_{N-1})$ implies, using (5), $\varphi_\Gamma(\omega_{N-1}) = 0$, which is justified only in (18).

*C. Location and characterization of a fault with reflection coefficient independent on the frequency*

In this case, since $\bar\Gamma$ is independent on the frequency, all the information given by $\bar H(\omega)$ is easily exploited, by solving the OLS problem:

$$\begin{bmatrix} \ln H(\omega_0) \\ \ln H(\omega_1) \\ \vdots \\ \ln H(\omega_{N-1}) \\ \varphi_H(\omega_0) \\ \varphi_H(\omega_1) \\ \vdots \\ \varphi_H(\omega_{N-1}) \end{bmatrix} = \begin{bmatrix} 1 & 0 & -2\alpha(\omega_0) \\ 1 & 0 & -2\alpha(\omega_1) \\ \vdots & \vdots & \vdots \\ 1 & 0 & -2\alpha(\omega_{N-1}) \\ 0 & 1 & -2\beta(\omega_0) \\ 0 & 1 & -2\beta(\omega_1) \\ \vdots & \vdots & \vdots \\ 0 & 1 & -2\beta(\omega_{N-1}) \end{bmatrix} \cdot \begin{bmatrix} \ln\Gamma \\ \varphi_\Gamma \\ l \end{bmatrix} \qquad (23)$$

This way of characterizing the fault is similar to that described by equation (13) in [22], but does not need a quadratic model for the coefficients $\alpha(\omega)$ and $\beta(\omega)$. This is, of course, only a special case of constraints $\bar\Gamma(\omega)$ that makes (4) and (5) a system of equations with a unique solution.

## V. TEST RESULTS ON SIMULATED CABLES

The aim of this Section is to evaluate, in typical cases, systematic errors of the proposed algorithms for SFWR. To this purpose, a set of tests have been performed on simulated cables, with exactly known parameters, and with noiseless SFWR signals. Errors associated to an imperfect knowledge of the cable or of the load parameters, and to the noise, are therefore eliminated. A complete uncertainty analysis, including e.g. the effect of noise, goes beyond the scope of the present work, and could be the subject of a separate paper.

Cables have been simulated using LineLab software [23], a MATLAB-based simulator of quasi-TEM transmission lines. It can simulate transmission lines with arbitrary dispersion models, and arbitrary profiles of the frequency-dependent primary parameters $R, C, G, L$. All simulations have been carried out mimicking an RG58-CU coaxial cable. A dispersive model for the cable primary parameters have been used [25], with geometrical and electrical parameters as shown in Fig. 6. The nominal characteristic impedance of the cable is $Z_0 = 50\,\Omega$, and this is also the exact internal impedance of the generator in the simulations. The actual characteristic impedance of the cable is slightly different and frequency-dependent, as for any real cable. For example, at the frequency $f = 1$ MHz, the true impedance is $|Z_0| \cong 49.5\,\Omega$, $\angle Z_0 \cong -0.12$ rad.

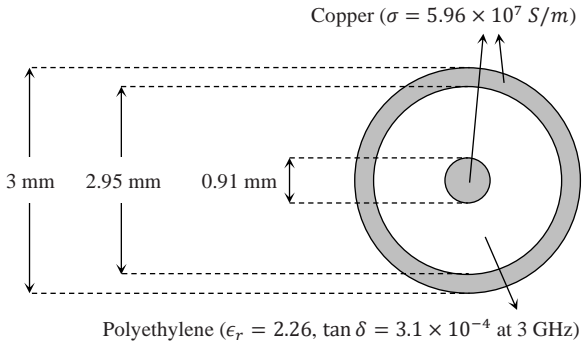

Fig. 6 Cross section and electrical parameters of the simulated cable.

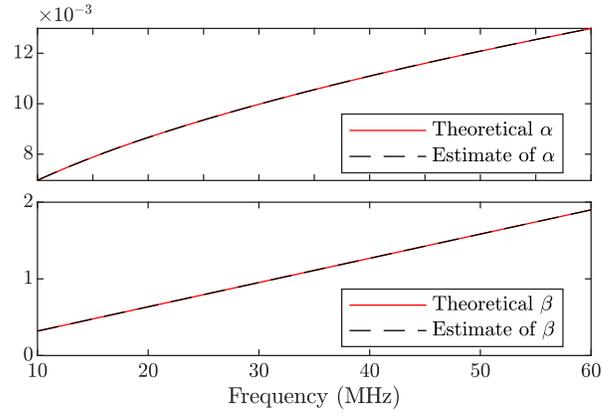

Fig. 8 Theoretical and estimated $\alpha(\omega_i)$ and $\beta(\omega_i)$ in the simulated reference cable.

Both series and shunt faults have been simulated like in [22]. The series impedance (Fig. 7a) may represent a defective point in the cable conductors (e.g. a junction), while the shunt impedance (Fig. 7b) may represent a damage in the dielectric of the coaxial cable. The reflection coefficient of the series and shunt faults are, respectively:

$$\bar{\Gamma}_S = \frac{Z_S}{Z_S + 2Z_0} \quad (24)$$

$$\bar{\Gamma}_P = -\frac{Z_0}{Z_0 + 2Z_P} \quad (25)$$

where $Z_S$ is the series impedance and $Z_P$ is the shunt impedance.

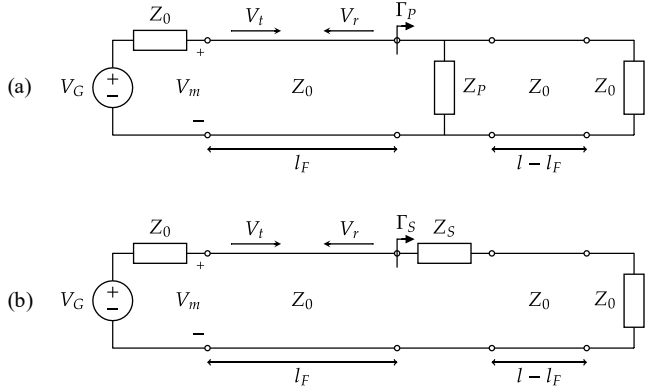

Fig. 7 Simulated cables with series (a) and shunt (b) fault.

Finally, in all the experiments the parameters of the SFWR signal are $\tau = 200$ ns, $T = 2$ μs, $f_0 = 10$ MHz, $N = 101$ frequencies, $\Delta f = 500$ kHz, $f_{N-1} = 60$ MHz. They have been chosen according to the criteria stated in Subsections II.B, III.C, IV.B.

### A. Characterization of a reference cable

In this simulation, the cable is 50 m long, without faults and with open termination ($l = 50$ m, $\Gamma = 1$). Functions $\hat{\alpha}(\omega_i)$ and $\hat{\beta}(\omega_i)$ have been obtained in the 101 test frequencies using (16), (17). Results are in Fig. 8. The maximum relative error is 0.10% for the estimate $\hat{\alpha}(\omega)$, and 0.02% for the estimate $\hat{\beta}(\omega)$.

### B. Fault location and characterization on a cable under test, with a generic reflection coefficient

In this simulation, the cable is 100 m long, and has a series capacitive fault at $l_F = 55$ m from the beginning of the line. The reflection coefficient at the fault is:

$$\bar{\Gamma}_S = \frac{Z_S}{Z_S + 2Z_0} = \frac{1}{1 + j\omega 2Z_0 C_F} \quad (26)$$

where $C_F = 100$ pF is the fault capacitance. The fault position has been estimated using (19) for each $\omega_i$, obtaining the results reported in Fig. 9.

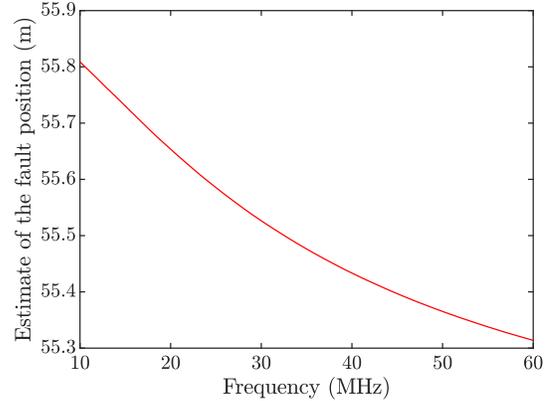

Fig. 9 Estimate at various frequencies of the position of a series capacitive fault (true $l_F = 55$ m). The smaller systematic error is achieved at the maximum frequency.

In accordance with (20), there is a systematic error decreasing at higher frequencies. At the highest frequency $f_{max} = 60$ MHz the estimate is $\hat{l} = 55.31$, with a relative error $e_r \cong 0.6\%$, as foreseen by (20).

The estimation of the magnitude $\Gamma(\omega)$ of the reflection coefficient is shown in Fig. 10. The maximum relative error over the theoretical value is 1.73%.

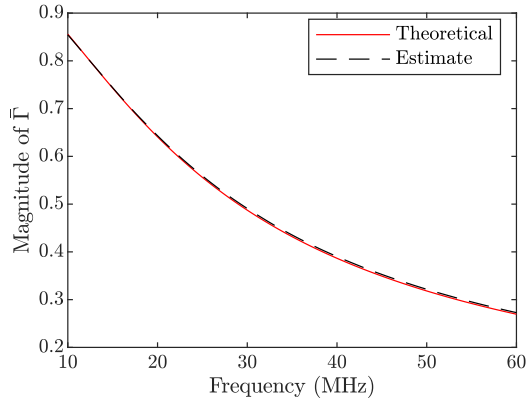

Fig. 10 Estimate of the magnitude $\Gamma(\omega)$ of the reflection coefficient

## C. Fault location and characterization on a cable under test, with a frequency-independent reflection coefficient

### 1) Reflection due to a mismatched load

For this case, a 100 m long cable terminated on a resistive load $Z_L$ have been simulated. For a resistive load, $\bar{\Gamma}(\omega)$ is independent on the frequency and it is possible to perform the simultaneous measurement of $l$ and $\bar{\Gamma}$ as described in Subsection IV.C. Seven values of $Z_L$ have been simulated, including short circuit, open circuit, and $Z_L = 50\ \Omega \cong Z_0$. In this limit case, the reflection coefficient is not exactly zero because the simulated cable has a "real" behavior, and $Z_0$ is not exactly 50 Ω.

Table II reports the estimations of $l$, Table III the estimations of $\bar{\Gamma}$.

TABLE II
ESTIMATED CABLE LENGTHS

| $Z_L$ (Ω) | $l$ (m) | $e_r(l)$ (%) |
|---|---|---|
| 0 | 99.98 | −0.0175 |
| 10 | 99.98 | −0.0196 |
| 30 | 99.97 | −0.0320 |
| 50 | 99.31 | −0.693 |
| 100 | 100.0 | $-1.28 \times 10^{-4}$ |
| 300 | 99.99 | −0.00606 |
| ∞ | 99.99 | −0.00750 |

Estimation errors are in general very low. Meaningful errors are observed only for the phase $\varphi_\Gamma$, when the load is near or equal to the characteristic impedance. For $Z_L = 50\ \Omega$ estimation errors get slightly larger (the error on $\varphi_\Gamma$ is not meaningful since $\Gamma \approx 0$), but the algorithms still achieve acceptable results.

TABLE III
ESTIMATED REFLECTION COEFFICIENT

| $Z_L$ (Ω) | $\Gamma$ | $e(\Gamma)$ | $\varphi_\Gamma$ (deg) | $e(\varphi_\Gamma)$ (deg) |
|---|---|---|---|---|
| 0 | 0.9999 | $-1.36 \times 10^{-4}$ | 180.0 | −0.00437 |
| 10 | 0.6506 | $-1.09 \times 10^{-4}$ | 179.5 | −0.218 |
| 30 | 0.2233 | $-1.21 \times 10^{-4}$ | 177.7 | −1.00 |
| 50 | 0.02748 | $-1.21 \times 10^{-3}$ | 23.50 | 13.0 |
| 100 | 0.3582 | $5.10 \times 10^{-6}$ | 1.299 | 0.569 |
| 300 | 0.7279 | $-6.48 \times 10^{-5}$ | 0.3395 | 0.146 |
| ∞ | 0.9999 | $-1.12 \times 10^{-4}$ | 0.004259 | −0.00426 |

### 2) Reflection due to a series or a parallel fault

In these simulations the reflections are due to point-like impedance discontinuities in the cable, as represented in Fig. 7. Mathematically, the situation is equivalent to that of a cable without faults, but with a mismatched load. Simulated fault impedances are purely resistive, so once again it is possible to perform measurements as described in Subsection IV.C.

The circuit was simulated considering a fault placed at $l_{F1} = 30$ m or $l_{F2} = 70$ m from the cable beginning with different values of impedances for each of the two fault types. Table IV, shows that the fault positions are estimation with very low systematic error in all the considered cases.

TABLE IV
ESTIMATED FAULT POSITIONS

| | $Z_F$ (Ω) | $l_{F1}$ (m) | $er(l_{F1})$ (%) | $l_{F2}$ (m) | $er(l_{F2})$ (%) |
|---|---|---|---|---|---|
| Series fault | 20 | 29.996 | −0.0130 | 69.990 | −0.0141 |
| | 50 | 29.995 | −0.0151 | 69.989 | −0.0150 |
| | 100 | 29.995 | −0.0172 | 69.989 | −0.0159 |
| | 200 | 29.994 | −0.0193 | 69.988 | −0.0168 |
| | 500 | 29.994 | −0.0213 | 69.988 | −0.0177 |
| Shunt fault | 5 | 30.002 | 0.00782 | 69.996 | −0.00517 |
| | 10 | 30.002 | 0.00631 | 69.996 | −0.00582 |
| | 20 | 30.001 | 0.00432 | 69.995 | −0.00668 |
| | 50 | 30.000 | 0.00159 | 69.995 | −0.00785 |
| | 200 | 30.000 | −0.00108 | 69.994 | −0.00899 |

Also, the systematic error in the estimation of the reflection coefficient $\bar{\Gamma}$ is very low: the maximum relative error affecting $\hat{\Gamma}$ is $3.67 \times 10^{-5}$, while the maximum error affecting $\hat{\varphi}_\Gamma$ is 0.42 degrees.

## VI. EXPERIMENTAL RESULTS

Effectiveness of the proposed algorithm was demonstrated with tests in real situations also.

### A. Experimental Setup

The experimental setup used in tests is depicted in Fig. 11. Both signal generation and acquisition are performed using a GW Instek MSO-2047EA oscilloscope, that is remotely controlled by a common personal computer. The output of the arbitrary waveform generator integrated in the GW Instek MSO-2047EA is connected to the input of the oscilloscope and to the cable under test through a T-connector. RG58-CU coaxial cables, of the same type of those used in simulations, were used in all experiments.

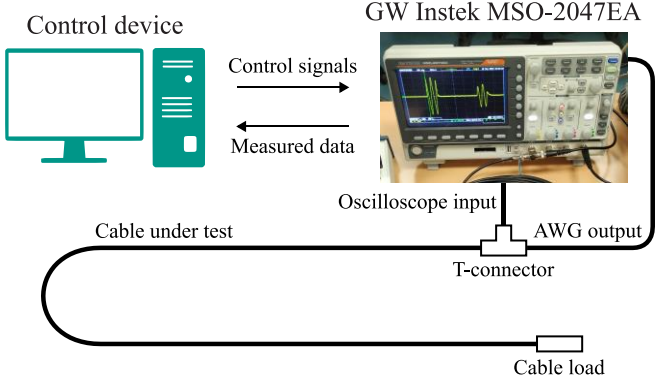

Fig. 11 Experimental setup

### B. Characterization of a reference cable

First, the cable was characterized as described in Subsection IV.A, using an open-ended 50 m long reference sample. The parameters of the SFWR signal were $\tau = 200$ ns, $T = 2$ µs, $f_0 = 10$ MHz, $N = 51$ frequencies, $\Delta f = 1$ MHz, $f_{N-1} = 60$ MHz. The propagation function obtained from the estimation procedure is depicted in Fig. 12.

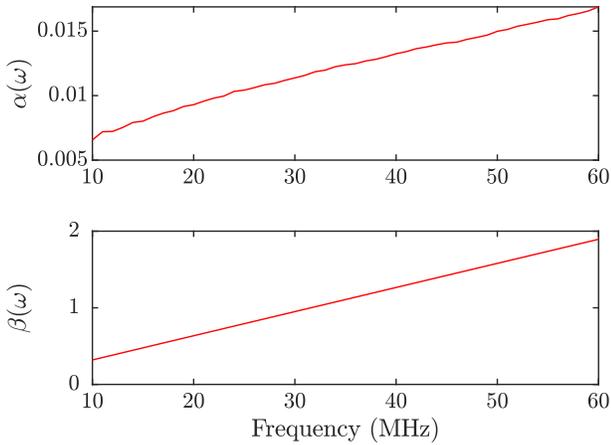

Fig. 12 Results of the characterization of the reference cable

### C. Fault location and characterization on a cable under test terminated on a resistive load

Then a 100 m long cable, terminated on different resistive loads, was analyzed. Some portions of the measured signal when the cable had a short-circuit termination are reported in Fig. 13. The reflection coefficient and the cable length were estimated as described in Subsection IV.C. Estimation results, reported in Table V, demonstrate the accuracy of the algorithm.

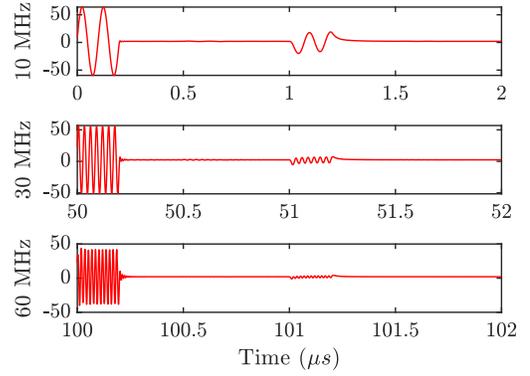

Fig. 13 Example of measured signal for a 100 m long coaxial cable terminated on a short circuit

TABLE V
ESTIMATED LENGTH AND REFLECTION COEFFICIENT

| $Z_L$ (Ω) (nominal value) | $\bar{\Gamma}$ (nominal value) | $\Gamma$ | $\varphi_\Gamma$ (deg) | $l$ (m) |
|---|---|---|---|---|
| 0 | -1 | 1.04 | 171.4 | 100.33 |
| 25 | -0.333 | 0.355 | 174.3 | 100.05 |
| 100 | 0.333 | 0.337 | -3.182 | 99.99 |
| ∞ | 1 | 1.04 | -4.841 | 100.01 |

## VII. CONCLUSIONS

The paper proposes a development of the interesting theory and methods presented in [22], to perform the reflectometry with sinusoidal bursts named SFWR. The main results in the paper are the following.

1) Formulae are given to choose the burst duration $\tau$, the separation between bursts $T$, and the frequency range $[\omega_{min}; \omega_{max}]$ of the bursts.

2) An algorithm is proposed to obtain the FRF $\bar{H}(\omega)$ at any frequency $\omega_i$ using a modified sinusoidal OLS fitting. The algorithm takes into account the fact that the response of $\bar{H}(\omega)$ to a sinusoidal burst contains transient terms, and is therefore a valid alternative to the time-frequency analysis employed in [22].

3) Algorithms are presented to extract the fault location $l$ and the fault reflection coefficient $\bar{\Gamma}(\omega)$ from the knowledge of $\bar{H}(\omega_i)$, even at a single frequency. These algorithms add up to those described in [22], which are perfectly valid, but requires some constraints (quadratic model of the propagation function, $\bar{\Gamma}$ independent on the frequency). Besides, the resolution in the measurement of $l$ is not linked to the number of frequencies.

4) Systematic errors associated to the method are determined by a simulation study, with cable and faults of realistic but exactly known characteristics. The errors are demonstrated to be negligible or very low.

By considering the results of this paper with that in [22], the feasibility of SFWR as a low-cost but very accurate reflectometric method is fully demonstrated. A complete uncertainty analysis is desirable, but can be the subject of a separate paper.

ACKNOWLEDGMENT

The authors wish to thank prof. Amerigo Trotta and Dr. Pietro Pappalardi for the useful hints and discussions in the initial stage of the work.